\documentclass[12pt]{article}
\usepackage{epsf}
\def\be{\begin{equation}}
\def\ee{\end{equation}}
\def\bea{\begin{eqnarray}}
\def\eea{\end{eqnarray}}

\begin{document}
\begin{titlepage}
\begin{center}
{\Large \bf William I. Fine Theoretical Physics Institute \\
University of Minnesota \\}
\end{center}
\vspace{0.2in}
\begin{flushright}
FTPI-MINN-17/12 \\
UMN-TH-3629/17 \\
July 2017 \\
\end{flushright}
\vspace{0.3in}
\begin{center}
{\Large \bf A mixing model for bottomoniumlike $Z_b$ resonances.
\\}
\vspace{0.2in}
{\bf  M.B. Voloshin  \\ }
William I. Fine Theoretical Physics Institute, University of
Minnesota,\\ Minneapolis, MN 55455, USA \\
School of Physics and Astronomy, University of Minnesota, Minneapolis, MN 55455, USA \\ and \\
Institute of Theoretical and Experimental Physics, Moscow, 117218, Russia
\\[0.2in]

\end{center}

\vspace{0.2in}

\begin{abstract}
The isovector bottomoniumlike exotic resonances $Z_b(10610)$ and $Z_b(10650)$ are very close to the thresholds of heavy meson pairs $B \bar B^* \, ( \bar B B^*)$ and $B^* \bar B^*$, and are naturally understood as `molecular' states made of the corresponding meson-antimeson pairs. The picture with a full separation of the two channels cannot be exact and a cross-feed between the states necessarily takes place. A simple model is considered here describing the mixing between the channels in terms of one parameter. The model reasonably describes the current data on the decays of the $Z_b$ resonances to bottomonium plus a pion and predicts the rate and a distinctive interference pattern for the decay $Z_b(10650) \to B \bar B^* \, ( \bar B B^*)$ as well as excess of the mass splitting between the resonances over the mass difference between $B^*$ and $B$ mesons.

\end{abstract}
\end{titlepage}

The twin bottomoniumlike resonances $Z_b(10610)$ and $Z_b(10650)$ found~\cite{bellez} by the Belle experiment in the decays $\Upsilon(5S) \to Z_b \pi$  necessarily contain a light quark-antiquark  pair in addition to the heavy $b \bar b$ pair since they come in full isotopic triplets with electrically charged, $Z_b^\pm$ and neutral~\cite{bellez0}, $Z_b^0$, states. Furthermore, their masses coincide within few MeV with the corresponding thresholds for pairs of heavy mesons, $B \bar B^*$ and $B^* \bar B^*$, and are thus interpreted~\cite{bgmmv} as `molecular'~\cite{ov} $S$ wave threshold states in the respective meson-antimeson channels with the quantum numbers $I^G(J^P)=1^+(1^+)$, which quantum numbers are also established by the experiment~\cite{belleqn,pdg}. 
The molecular picture is strongly favored by the data~\footnote{It should be mentioned that alternative descriptions of the $Z_b$ resonances are discussed in the literature, in particular based on a diquark-antidiquark model~\cite{ahw,ampr,epp}. A discussion can be found in the review \cite{ghmwzz}.}. In particular, it explains the apparent breaking of the heavy quark spin symmetry (HQSS) in the processes with the $Z_b$ resonances. Namely, these resonances decay to the states of bottomonium plus pion, with a comparable rate for the bottomonium $b \bar b$ pair being in the ortho- spin state ($S_{b \bar b} = 1$), $Z_b \to \Upsilon(nS) \, \pi$, $n=1,2,3$, and in the para- spin state ($S_{b \bar b} = 0$), $Z_b \to h_b(kP) \, \pi$, $k=1,2$. In the $B \bar B^* \, ( \bar B B^*)$ and $B^* \bar B^*$ pairs the spin of the $b$ quark is correlated with that of the light antiquark $\bar q$ in the meson, while the spin of $\bar b$ is correlated with that of $q$. As a result the spin state of the $b \bar b$ pair in a molecular state is generally mixed. In particular, for the spin structure of the relevant $S$ wave states of the meson-antimeson pairs one finds~\cite{bgmmv} in terms of the total spin of the $b \bar b$ and $q \bar q$ pairs 
 \bea
&&Z_b \, \sim  \,\left | B^* \bar B, B \bar B^* \right \rangle_{I^G(J^P)=1^+(1^+)} \, \sim \, {1 \over \sqrt{2}} \left ( 0^-_{b \bar b}\otimes 1^-_{q \bar q} - 1^-_{b \bar b}\otimes 0^-_{q \bar q} \right )~, \nonumber \\
&&Z'_b \, \sim \, \left | B^* \bar B^*\right \rangle_{I^G(J^P)=1^+(1^+)} \, \sim \, {1 \over \sqrt{2}} \left ( 0^-_{b \bar b}\otimes 1^-_{q \bar q} + 1^-_{b \bar b}\otimes 0^-_{q \bar q} \right )~,
\label{zspin}
\eea 
which explains the presence of $b \bar b$ states with both possible values of the total spin in the decay products of the resonances, if the states $Z_b$ and $Z'_b$ are identified as the observed peaks $Z_b(10610)$ and $Z_b(10650)$. The purpose of the present paper is to consider a deviation from the ideal mixing structure described by these relations and to discuss a  model where all such deviation is parametrized in terms of one mixing angle $\theta$:
\bea
&& Z_b(10610) = \cos \theta \, Z_b - \sin \theta \, Z'_b~, \nonumber \\
&& Z_b(10650) = \sin \theta \, Z_b + \cos \theta \, Z'_b~.
\label{mix}
\eea
Such simplified approach is well known to be helpful in  discussion of e.g. the isospin violation in terms of $\rho - \omega$ mixing, or of the flavor SU(3) violation, $\eta - \eta'$ and $\omega - \phi$ mixing. This simple mixing description is certainly an approximation, since the amount of mixing between the spin states is likely to be a function of other variables in the wave functions of the states, e.g. of the distance. In other terms, there may be many more states involved in the mixing, such as e.g. the continuum of the heavy meson pairs, $B \bar B^* \, ( \bar B B^*)$ and $B^* \bar B^*$, with the amount of the mixing depending on the excitation energy. Generally, the scattering dynamics in coupled channels is determined by interaction between mesons and involves more parameters (a somewhat general discussion can be found in e.g. Refs.~\cite{mp,Guoetal} and the recent review~\cite{ghmwzz}). The discussed approximation in terms of one overall mixing angle is applicable if the near-threshold dynamics in the $I^G(J^P)=1^+(1^+)$ meson-antimeson channels is strongly dominated by the $Z_b$ resonances. This may indeed be the case since the observed~\cite{bellebb} spectra in the processes $\Upsilon(5S) \to B^{(*)} \bar B^{(*)} \pi$ are apparently fully given by the $Z_b$ resonances with no significant nonresonant contribution. 

It should be mentioned that a similar model based on mixing of just two states might be applicable to the charmoniumlike $Z_c(3900)$ and $Z_c(4020)$ resonances at the respective $D \bar D^*$ and $D^* \bar D^*$ thresholds. However, the data on the properties of these states to charmonium plus pion and also on the behavior in the $D \bar D^* (\bar D D^*)$ channel near the $Z_c(4020)$ peak are currently insufficient to draw a conclusion on the relevance of the discussed model in the charmonium sector.

The spin structure in Eq.(\ref{zspin}) is that of  free non-interacting meson pairs and it would be preserved if the interaction between the mesons did not depend on the spin of either heavy or light quark-antiquark pair~\cite{mv16}. The suppression of the dependence on the spin of the heavy quarks is equivalent to HQSS. On the other hand generally there is no light quark spin symmetry (LQSS), and one would expect deviations from the `ideal' spin structure given by Eq.(\ref{zspin}) and other predictions from such symmetry. One of the consequences of the ideal spin structure is absence of decays of the heavier state $Z'_b$ to the lighter meson pairs $B \bar B^* \, ( \bar B B^*)$. Indeed, the only reason for suppression of this decay, fully allowed otherwise, is the spin orthogonality of the states in Eq.(\ref{zspin}). Furthermore, in the limit of the spin independence of the interactions the $Z_b$ and $Z'_b$ are dynamically the same state, one made of $B \bar B^* \, ( \bar B B^*)$ and the other of $B \bar B^*$, merely shifted in mass by $\Delta= M(B^*) - M(B)$. Thus one should expect the relation $M(Z'_b) - M(Z_b) = \Delta$, and that the partial decay rates as well as the total widths of the two resonances should be the same, modulo a difference in the phase space caused by the mass difference. Some of these predictions are close to the experimentally observed properties and some are less so, or unknown. In particular,  an analysis~\cite{bellebb} of the spectra of the  $B^* \bar B \,  (B \bar B^*)$ and $B^* \bar B^*$ pairs in the decays $\Upsilon(5S) \to B^{(*)} \bar B^{(*)} \, \pi$ shows no significant features (above the uncertainties) in the spectrum of invariant mass in the channel  $B^* \bar B + B \bar B^*$ at the mass of $Z_b(10650)$ that would indicate a presence of a coupling of this channel to the latter higher resonance.  The current uncertainty in the splitting between the measured values of the masses~\cite{pdg}  $10652.2 \pm 1.5$\,MeV and $10607.2 \pm 2.0$\,MeV is too large to see a possible deviation from the meson mass splitting $\Delta=45.18 \pm 0.23$\,MeV. The measured total widths of the two peaks, $\Gamma[Z_b(10610)]= 18.4 \pm 2.4$\,MeV and $\Gamma[Z_b(10650)]= 11.5 \pm 2.2$\,MeV, also suffer from a considerable uncertainty preventing one from concluding whether considering them approximately equal would be a good starting approximation. The most apparent indication of a breaking of the ideal symmetry in the $Z_b$ resonances is provided by the observed pattern of the relative, between the two resonances, decay rates to ortho- and para- bottomonium. Namely, the data~\cite{bellebb,pdg} strongly suggest that the decays of the heavier resonance $Z_b(10650)$ to the ortho- bottomonium states, $\Upsilon(nS) \, \pi$ are somewhat weaker than the corresponding decays of the lighter $Z_b(10610)$, while in the decays to the para- states, $h_b(kP) \, \pi$ the relative yield from the two peaks is reversed.  A violation of the ideal symmetry limit should not come as a surprise, since neither a LQSS can be justified in QCD, nor a separation of the spin structures in Eq.(\ref{zspin}) is likely to be sustainable for reasons based on unitarity. Indeed, the states $Z_b$ and $Z'_b$ have common decay channels, and can thus mix with each other, and the absorptive part of the mixing through a given (on-shell) channel $X$ can be estimated as
\be
\theta_X \sim { 1 \over \Delta} \sqrt{ \Gamma(Z_b \to X) \Gamma (Z'_b \to X) }.
\label{tx}
\ee
Using the data~\cite{bellebb,pdg} for the decays of the $Z_b$ resonances to bottomonium plus pion, one readily finds that the largest contributing decay channel is $h_b(2P) \, \pi$, giving $\theta_{h_b(2P) \pi} \approx  3 \times 10^{-2}$, with other intermediate channels contributing significantly less. In what follows it will be argued that the current data suggest that the mixing angle is significantly larger (although can still be considered as small), $\theta \approx 0.2$, so that the mixing arises dominantly from off-shell intermediate states and its absorptive part can be neglected.

Clearly, a mixing described by Eq.(\ref{mix}) tilts the $b \bar b$ spin structure in the resonances, so that at a small positive $\theta$ the resonance $Z_b(10610)$ gets a larger $1^-_{b \bar b}$ spin component, while the $0^-_{b \bar b}$ para- component is enhanced in the $Z_b(10650)$ resonance, which qualitatively agrees with the data on the relative rates of decays to  $\Upsilon(nS) \, \pi$ and $h_b(kP) \, \pi$. In order to take into account the kinematical differences between the decays from the two resonances one can write the decay amplitudes according to the parity and the current algebra requirements~\cite{bgmmv}:
\bea
A[ Z_b \to \Upsilon(nS) \pi] & = & C[Z_b \to \Upsilon(nS) \pi] \, (\vec Z_b \cdot \vec \Upsilon) \, E_\pi~, \nonumber \\ 
A[Z_b \to h_b(kP) \pi] & = & D[Z_b \to h_b(kP) \pi] \, ( [\vec Z_b \times \vec h_b] \cdot \vec p_\pi)~,
\eea
where $\vec Z_b$, $\vec \Upsilon$ and $\vec h_b$ are the polarization amplitudes of the spin-1 resonances, $\vec p_\pi$ is the pion momentum and $E_\pi$ is its energy. One can then expect that the coefficients $C$ and $D$ are largely free of kinematical differences between $Z_b$ and $Z'_b$ resonances, and in fact that in the limit of the ideal symmetry the coefficient for each particular decay amplitude is the same between the $Z_b$ and $Z'_b$ resonances with the relative sign given by that of the corresponding spin component in Eq.(\ref{zspin}). Then in the first order in the mixing in Eq.(\ref{mix}) the ratio of the coupling strengths of the $Z_b(10610)$ and $Z_b(10650)$ states is found as
\be
 { \Gamma[Z_b(10610) \to \Upsilon(nS) \pi]/(E_\pi^2 p_\pi) \over  \Gamma[Z_b(10650) \to \Upsilon(nS) \pi]/(E_\pi^2 p_\pi)}  = \left ( {\cos \theta + \sin \theta \over \cos \theta - \sin \theta } \right )^2
\label{crat}
\ee
for each $n$, and 
\be
 { \Gamma[Z_b(10650) \to h_b(kP) \pi]/p_\pi^3 \over  \Gamma[Z_b(10610) \to h_b(kP) \pi]/p_\pi^3}   = \left ( {\cos \theta + \sin \theta \over \cos \theta - \sin \theta } \right )^2
\label{drat}
\ee
for each $k$. 

The data~\cite{bellebb,pdg} correspond to the values of the ratios in Eq.(\ref{crat}) $3.6 \pm 1.8$, $3.2 \pm 1.4$ and $2.3 \pm 1.15$ for $n =1,$ 2 and 3 respectively in units of $\Gamma_{tot} [Z_b(10610)]/\Gamma_{tot}[Z_b(10650)]$, and to those in Eq.(\ref{drat}) equal to $2.0 \pm 0.9$ at $k=1$ and $2.2 \pm 0.9$ at $k=2$ in units of $\Gamma_{tot} [Z_b(10650)]/\Gamma_{tot}[Z_b(10610)]$. Assuming that the total widths are the same, a  fit to these data gives approximately $\theta \approx 0.2 \pm 0.1$. 

Within the discussed approach, and in the first order in the mixing, the total decay rates of the physical resonances are expected to be the same, which agrees with the data only if the experimental errors are taken into account. One can attempt to allow for a different overall decay rate between the resonances by assuming that the rates of all the decays of the lower resonance $Z_b(10610)$ are enhanced by a common factor $F$. [In other words, the r.h.s. in Eq.(\ref{crat}) is multiplied by $F$, while that in Eq.(\ref{drat}) receives the factor $1/F$.] Then a fit to the data with two parameters $F$ and $\theta$ produces the central value $F = 1.17$ and an essentially the same central value for the mixing angle $\theta$. A one-sigma contour for the two parameter fit is shown in Fig.~1. 

\begin{figure}[ht]
\begin{center}
 \leavevmode
    \epsfxsize=8cm
    \epsfbox{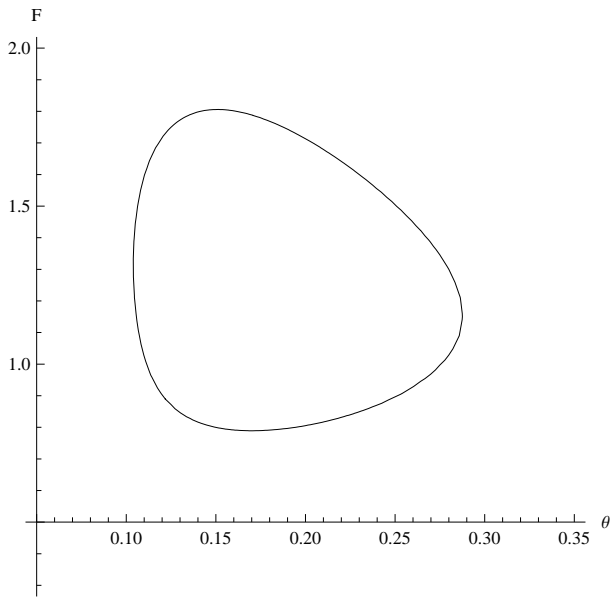}
    \caption{The one-sigma contour for the fit to the available data of the mixing angle $\theta$ and the relative enhancement factor $F$ for the decay rates of $Z_b(10610)$.  }
\end{center}
\end{figure}

The fitted value of the mixing angle provides an estimate for the scale of the (yet to be observed) decay $Z_b(10650) \to B \bar B^* (\bar B B^*)$: $\Gamma[Z_b(10650) \to B \bar B^* (\bar B B^*)]/\Gamma[Z_b(10650) \to B^* \bar B^*] \sim \theta^2 \sim 0.04$. This result however requires further specification, given the kinematical differences in the compared decay channels in a specific experimental setting. Namely the yield of the heavy meson pairs at the $Z_b$ resonances is observed~\cite{bellebb} by studying the final states $B \bar B^* \pi \,  (\bar B B^* \pi)$ and $B^* \bar B^* \pi$ at the energy of the $\Upsilon(5S)$ resonance in $e^+e^-$ annihilation. There however may potentially be a tension between the discussed picture of the mixing and the $\Upsilon(5S)$ data. Namely, the relative strength of the coupling of this resonance to the channels $Z_b(10610) \, \pi$ and $Z_b(10650) \, \pi$ comes out approximately equal after taking into account the kinematical differences~\cite{mv16_2}. On the other hand, if the $\Upsilon(5S)$ was a  $J^{PC}=1^{--}$ state of a pure $b \bar b$ (ortho-) quark pair, its coupling to $Z_b(10610) \, \pi$ would be enhanced relative to $Z_b(10650)$ by the mixing in the way described by Eq.(\ref{crat}). However, unlike for the pure bottomonium states, the $b \bar b$ spin structure of the $\Upsilon(5S)$ is not protected and in fact can be modified by the near threshold enhancement of the HQSS breaking~\cite{mv12}, e.g. by mixing with $P$ wave meson-antimeson pairs. The observed sign of the interference between the $Z_b(10610)$ and $Z_b(10650)$ resonances in the processes $\Upsilon(5S) \to \Upsilon(nS) \pi \pi$ and $\Upsilon(5S) \to h_b(kP) \pi \pi$ suggests that even though the absolute values of the amplitudes can be modified by the possible HQSS breaking in $\Upsilon(5S)$, the relative signs of the amplitudes still agree with those for a pure $1^-_{b \bar b}$ spin state.  Within the sign convention used in Eq.(\ref{zspin}) this corresponds to an opposite sign of the coupling of the $\Upsilon(5S)$ to $Z_b \pi$ and $Z'_b \pi$. The spectrum of the invariant mass of the $B \bar B^* (\bar B B^*)$ pairs with the interference between the two $Z_b$ resonances is then given as
\be
{d \sigma \over d W} \propto \left | {1 \over W - M_1 + i \, \Gamma_1/2} - { \theta \over W- M_2 + i \, \Gamma_2/2} \right |^2 \, p \, E_\pi^2 \, p_\pi~,
\label{sbb}
\ee
where $M_1$ and $M_2$ ($\Gamma_1$ and $\Gamma_2$) are the masses (widths) of the $Z_b(10610)$ and $Z_b(10650)$ resonances and $p$ is the c.m. momentum of each of the heavy mesons. The interference behavior described by this expression is illustrated in Fig.~2. 

\begin{figure}[ht]
\begin{center}
 \leavevmode
    \epsfxsize=12cm
    \epsfbox{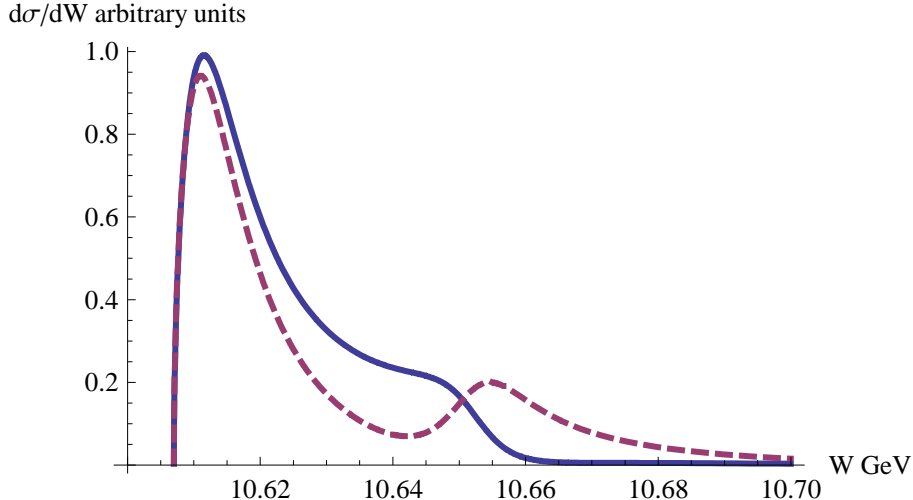}
    \caption{The expected spectrum of the invariant mass $W$ of the $B \bar B^* (\bar B B^*)$ pairs across the $Z_b(10610)$ and $Z_b(10650)$ in the discussed mixing model with $\theta=0.2$ (solid). Also shown for illustration is the spectrum (dashed) for the case where the sign of the mixing angle is opposite.}
\end{center}
\end{figure}

The discussed two-state mixing scheme is well known to produce a definite mass shift of the eigenstates: the splitting $\Delta$ is increased by $2 \theta^2 \, \Delta$, so that one can expect the mass difference between the $Z_b(10650)$ and $Z_b(10610)$ to be 
\be
M_2 -M_1 = (1+ 2 \theta^2) \, \Delta~.
\label{mdif}
\ee
At $\theta \approx 0.2$ this yields approximately $48.8$\,MeV. Given the uncertainty in $\theta$ this specific number may change. However the mass difference should be necessarily larger than $\Delta$ if the simple mixing model is of any relevance to the discussed resonances. In a general potential model with a mixing potential between the $B \bar B^*$ and $B^* \bar B^*$ channels the sign of the shift of the mass splitting from $\Delta$ is not fixed and depends on the details of the potential. However, as previously mentioned, the current experimental errors in the masses are yet too large for testing predictions of models.

In summary. The current status of deviation from the ideal spin structure in Eq.(\ref{zspin}) is not clear due to large experimental uncertainties. Some data, in particular on the relative strength of pion transitions from the $Z_b(10610)$ and $Z_b(10650)$ resonances, suggest effects of such deviation, while the apparent suppression of the coupling of the heavier state $Z_b(10650)$ to the lighter meson channel $B \bar B^* (\bar B B^*)$ indicates that the deviation is rather small. The separation of the states in Eq.(\ref{zspin}) can not be exact, e.g. due to existence of common decay channels (with no apparent cancellation between them). It is discussed here that the current data can be reconciled within a simple model of mixing of two states described by one angle as given by Eq.(\ref{mix}), with the value of the angle $\theta \approx 0.2$. This model then predicts a definite interference pattern in the process $\Upsilon(5S) \to B \bar B^* (\bar B B^*) \, \pi$ and the mass splitting between the $Z_b$ resonances that should be larger than the difference between the masses of $B^*$ and $B$ mesons.

This work is supported in part by U.S. Department of Energy Grant No.\ DE-SC0011842.

\end{document}